\documentclass[nohyperref]{article}

\usepackage{microtype}
\usepackage{graphicx}
\usepackage{subfigure}
\usepackage{booktabs} 

\usepackage{hyperref}

\usepackage[accepted]{icml2022}

\usepackage{amsmath}
\usepackage{amssymb}
\usepackage{mathtools}
\usepackage{amsthm}

\usepackage[capitalize,noabbrev]{cleveref}

\theoremstyle{plain}

\theoremstyle{definition}

\theoremstyle{remark}

\usepackage[textsize=tiny]{todonotes}

\icmltitlerunning{Conformal Prediction for Mammographic Breast Density Rating}

\begin{document}

\twocolumn[
\icmltitle{Three Applications of Conformal Prediction for Rating Breast Density in Mammography}




\begin{icmlauthorlist}
\icmlauthor{Charles Lu}{a}
\icmlauthor{Ken Chang}{a,b}
\icmlauthor{Praveer Singh}{a,d}
\icmlauthor{Jayashree Kalpathy-Cramer}{a,c,d}
\end{icmlauthorlist}

\icmlaffiliation{a}{Athinoula A. Martinos Center for Biomedical Imaging, Massachusetts General Hospital, Charlestown, MA, USA}
\icmlaffiliation{b}{Memorial Sloan Kettering Cancer Center, New York, NY, USA}
\icmlaffiliation{c}{Department of Radiology, Harvard Medical School, Boston, MA, USA}
\icmlaffiliation{d}{Department of Ophthalmology, University of Colorado School of Medicine, Aurora, CO, USA}

\icmlcorrespondingauthor{Charles Lu}{clu@mgh.harvard.edu}

\icmlkeywords{Machine Learning, ICML}

\vskip 0.3in
]



\printAffiliationsAndNotice{}  

\begin{abstract}
    Breast cancer is the most common cancers and early detection from mammography screening is crucial in improving patient outcomes. 
    Assessing mammographic breast density is clinically important as the denser breasts have higher risk and are more likely to occlude tumors. 
    Manual assessment by experts is both time-consuming and subject to inter-rater variability. 
    As such, there has been increased interest in the development of deep learning methods for mammographic breast density assessment. 
    Despite deep learning having demonstrated impressive performance in several prediction tasks for applications in mammography, clinical deployment of deep learning systems in still relatively rare; historically, mammography Computer-Aided Diagnoses (CAD) have over-promised and failed to deliver. 
    This is in part due to the inability to intuitively quantify uncertainty of the algorithm for the clinician, which would greatly enhance usability.
    Conformal prediction is well suited to increase reliably and trust in deep learning tools but they lack realistic evaluations on medical datasets. 
    In this paper, we present a detailed analysis of three possible applications of conformal prediction applied to medical imaging tasks: distribution shift characterization, prediction quality improvement, and subgroup fairness analysis. 
    Our results show the potential of distribution-free uncertainty quantification techniques to enhance trust on AI algorithms and expedite their translation to usage.
\end{abstract}

\section{Introduction}\label{intro}
    Deep learning (DL) methods have demonstrated superior performance for tasks in several application areas of medicine including radiology, pathology, electronic health records, drug discovery, and hospital operations \cite{topol2019high}. 
    Despite the high promise, the deployment of DL algorithms has been lagging. 
    Surrounding work in medical DL has highlighted a host of challenges towards clinical utility which importantly include trustworthiness, fairness, and robustness. Despite the high overall performance of automated algorithms, there are varied scenarios under which these algorithms fair poorly on yielding trustworthy clinical explainability~\cite{assessing_saliency}. 
    Given the high-stakes decisions made within medical practice, the inability to know when to trust or not trust the output of an algorithm is a major impediment to clinical adoption~\cite{10.1038/s42256-018-0004-1}.
    Toward the goal of ensuring more reliable DL-based software medical devices, 
    uncertainty quantification techniques have been proposed to provide estimates of prediction performance, model failure, and out-of-distribution generalization.
    Conformal prediction has especially emerged as a well-suited framework to provide distribution-free and intuitive uncertainty quantification technique for medical DL~\cite{10.1038/s41746-020-00367-3}.
    
    In this paper, we examine three clinically useful applications of conformal prediction on a multi-institutional screening mammography dataset for the task of automated breast density assessment.
    
        \begin{figure*}[ht]
        \begin{center}
        \centerline{\includegraphics[width=\textwidth]{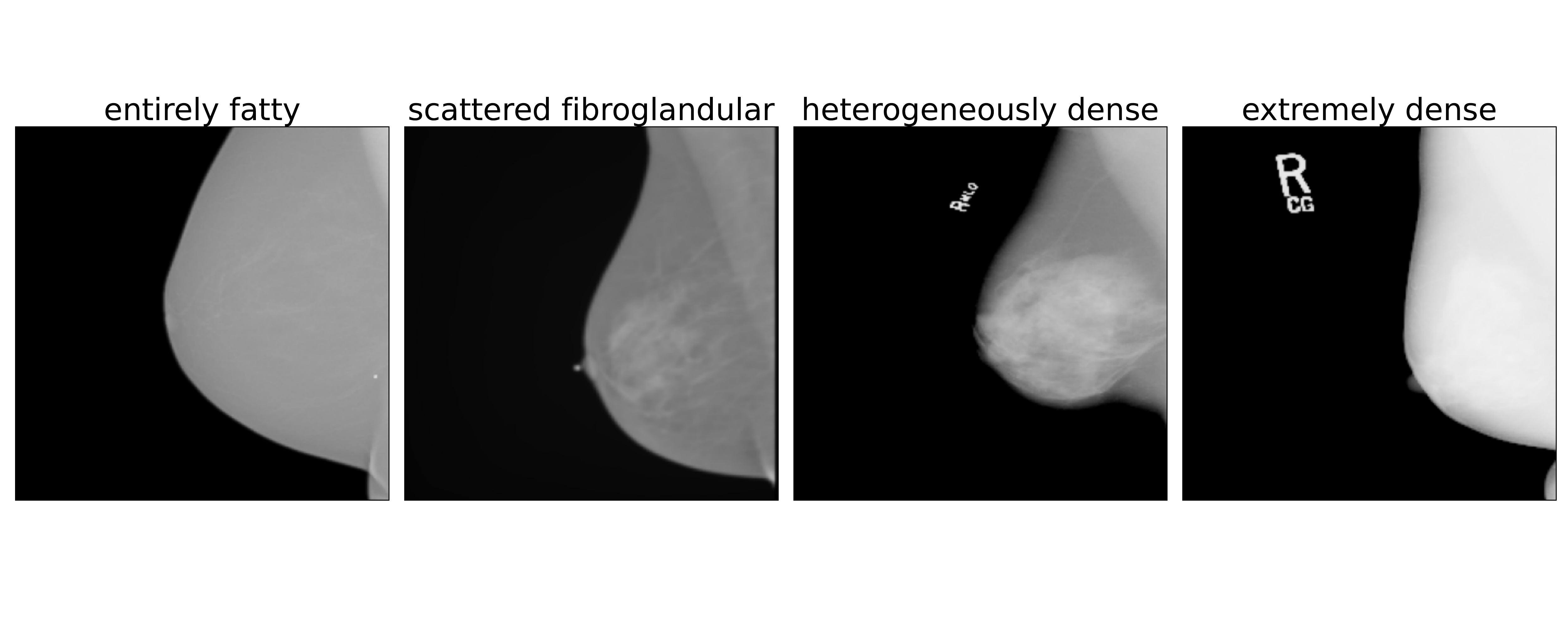}}
        \vskip -0.6in
        \caption{
        Accurate assessment of breast density, as measured by relative amounts of fibroglandular tissue, is important for the interpretation of mammographies, and is commonly rated according to BI-RADS criteria: entirely fatty, scattered, heterogeneously dense, and extremely dense.
        }
        \label{fig:birads}
        \end{center}
        \end{figure*}
        
\section{Automated Breast Density Rating in Mammography}\label{breast-density}
    Breast cancer is the most common invasive cancer in women with an estimated 2.3 million new cases diagnosed annually~\cite{ukasiewicz2021BreastCR}.
    Due to the high incidence of breast cancer among the general population, mammography imaging and breast examinations have become routine practice for healthy women, such as the United States, which recommend screening every two years between the ages of 50 and 74~\cite{siu2016screening}. 
    
    Due to the high volume of mammography examinations performed every year, Computer-Aided Diagnosis (CAD) tools have a high potential to augment or even automate several aspects of typical radiological workflows for breast cancer screening. This is key from a workflow efficiency standpoint as well as a way to address the inter-rater variability among experts with manual assessment \cite{sprague2016variation}.
    While previous attempts to deploy CAD in mammography have been fraught with unexpected difficulties and overstated expectations~\cite{10.1016/j.jacr.2017.12.029,Nishikawa2018ImportanceOB}, many expect deep learning-based CAD to provide substantial benefit and improved efficiency of breast cancer detection in mammography based on several promising studies~\cite{Wang2016DiscriminationOB,10.1038/s41598-019-48995-4,doi:10.1148/radiol.2018180694,doi:10.1148/radiol.2019182716,Kyono2019ImprovingWE,McKinney2020InternationalEO}.
    
    Besides detection of malignant tumors, assessment of breast density, as measured by fibroglandular tissue, is crucial to correct mammographic interpretation and as denser breasts are more likely to occlude cancers and are associated with increased risk~\cite{doi:10.1148/rg.352140106}.
    From a clinical standpoint, patients identified with high breast density may benefit from supplemental imaging examination such as ultrasound or MRI.
    
    Previous studies have shown that deep learning methods are effective for automatic breast density rating in mammography ~\cite{8462671,doi:10.1148/radiol.2018180694,10.1016/j.jacr.2020.05.015}. Importantly, mammography, as with many other imaging modalities, are prone to distribution shifts which can negatively impact performance such as different scanners and patient populations \cite{yala2021toward}. Being able to identify uncertainty within the context of distribution shift is key.
    
    Additionally, breast densty has been shown to be correlated with patient age, race, and breast volume (as measured by cup size)~\cite{10.1001/jamanetworkopen.2021.22810,10.1093/jnci/djw104,doi:10.1186/bcr2942.Epub}. A recent study has also show that patient race can be directly identified from mammography \cite{gichoya2022ai}. In light of various studies showing that medical DL algorithms can be biased across different patient subgroups \cite{seyyed2021underdiagnosis,larrazabal2020gender}, it is crucial to investigate uncertainty in the lens of fairness as well. 
    
        \begin{figure*}[ht]
        \vskip 0.2in
        \begin{center}
        \centerline{\includegraphics[width=0.9\textwidth]{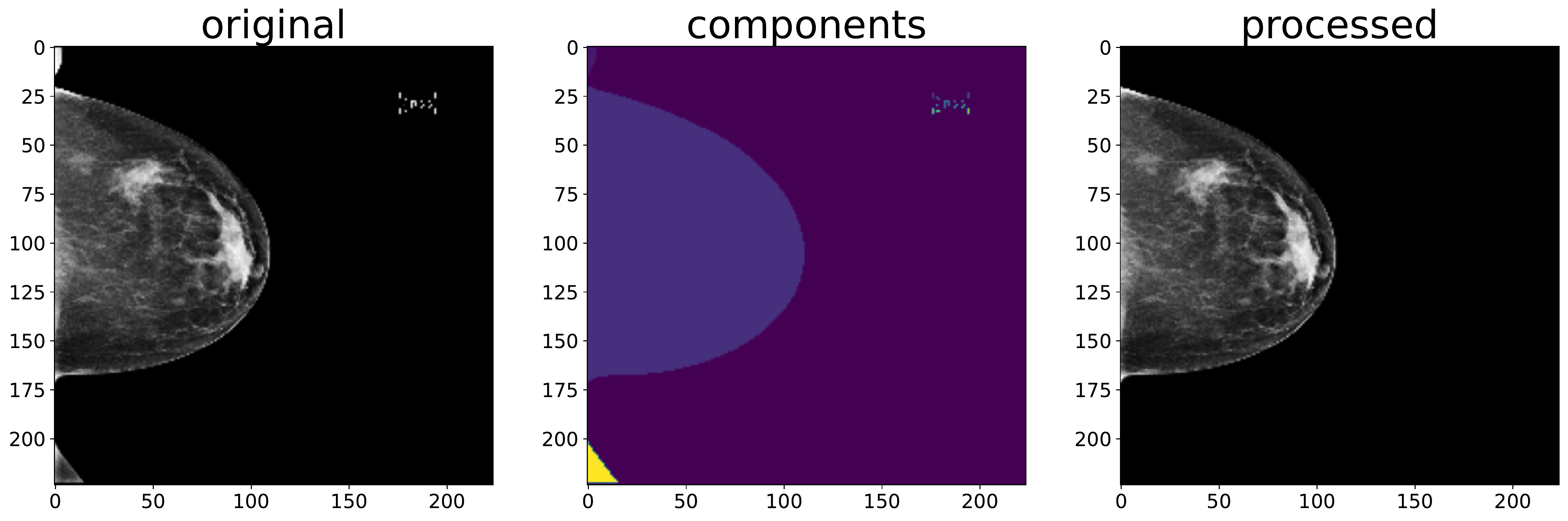}}
        \caption{Processed was done to remove confounding imaging artifacts and quantify breast size.}
        \label{fig:process}
        \end{center}
        \vskip -0.2in
        \end{figure*}
    
\section{Conformal Predictions}\label{conformal-prediction}
    Pioneered by ~\citet{10.5555/1062391}, conformal prediction is a novel approach to distribution-free uncertainty quantification relying on the notion of a ``nonconformality score'' which measures the distance of a new data point from previously seen points.
    If $S: \mathcal{X} \rightarrow \Delta^{\lvert\mathcal{Y}\rvert}$ are the softmax scores from some classifier model $f: \mathcal{X} \rightarrow \mathcal{Y}$, then an example nonconformality score could be
    \begin{equation}
        \small
        \text{NC}(x, y) := 1 - S(x)_y,
    \end{equation}
    where $y$ is the index of the ground truth label. 
    The scores of new data points can then be calculated and compared to the empirical distribution of scores calibrated on a held-out dataset.
    For an example classification task, the 90-th quantile of nonconformality scores can be estimated on a validation set and prediction sets can be formed for each test point by taking those classes that have softmax scores above this quantile threshold.
    
    Crucially, this form of inductive conformal prediction (also known as split conformal prediction) makes no assumptions on the underlying model or dataset and have been utilized for black box models such as deep learning to provide statistical guarantees of marginal coverage, defined as the average probability that the true class will be contained in the prediction set~\cite{4410411,10.1080/01621459.2012.751873,Sadinle_2018,10.5555/3495724.3496026,angelopoulosraps}
    
    Mathematically, marginal coverage can be expressed as the following:
    \begin{equation}
        \small
        1 - \alpha \; \leq \; P\left(y_i \; \in \; \left\{j \in \mathcal{Y} \mid S(x_i)_j > (1 - \hat{q}) \right\} \right),
    \end{equation}
    where $\alpha$ is a preset miscoverage level, $\{(x_i, y_i)\}_{i=1}^n$ are labeled test points, and $\hat{q}$ is the empirical $1-\alpha$ quantile of nonconformality scores with small finite sample correction:
    \begin{equation}
        \small
        \hat{q} = \text{quantile}\left(\{\text{NC}(x_i,y_i) \mid (x_i, y_i) \in D_\text{cal} \}, \frac{\lceil (1-\alpha) (m+1) \rceil}{m}\right)
    \end{equation}
    that is estimated on a calibration set, $D_\text{cal} := \{(x_i, y_i)\}_{i=1}^m$, that is assumed to be drawn exchangeably with the test set.
    
    Conformal prediction can be especially suited for medical domain to provide meaningful notions of uncertainty to aid clinical decision-making~\cite{10.1038/s41746-021-00504-6,https://doi.org/10.48550/arxiv.2109.04392,kompa,10.1007/s41666-021-00113-8}.
    Compared to other uncertainty and calibration techniques, which often assume direct access to the model (e.g. modification to loss function during training, scaling logits using original data distribution) that may not be practical for third-party medical devices.
    
\section{Experiments}\label{experiments}
    \subsection{Data and Implementation details}
        We use the Digital Mammographic Imaging Screening Trial (DMIST) dataset, which contains 108,230 mammograms collected from 33 sites that were interpreted by a total of 92 radiologists~\cite{Pisano2005DiagnosticScreening}. 
        We partition this dataset into 60\% training, 10\% validation, and 30\% testing sets.
        Additionally, we use an external testing dataset of 8,603 mammograms collected from Massachusetts General Hospital from 2010 following IRB approval.
        
        Breast density was assessed according to BI-RADS criteria into one of four categories: ``entirely fatty'', ``scattered fibroglandular'', ``heterogeneously dense'', or ``extremely dense''~\cite{bi-rads}.
    
        To assess the performance of breast density rating task, we use linearly weighted Cohen's Kappa score, $\kappa \in [-1, 1]$, which is commonly used to measure agreement between two raters, where $P_o$ is the proportion of observed agreement and $P_e$ is the agreement due to random chance.
        
        \begin{equation}
            \kappa = \frac{P_o - P_e}{1 - P_e}
        \end{equation}
        
        To avoid confounding bias from burnt in imaging artifacts, we process each image by taking connected components and keeping only the largest component (see Figure\ref{fig:process}).
        Additionally, this processing allows a simple quantification of breast volume to be the sum of non-zero pixels in the image.
        We scale pixel intensities to $[0, 1]$, apply random vertical flipping during training for data augmentation, and repeat across the channel dimension $3$ times in order to initialize each model with pretrained ImageNet weights~\cite{deng2009imagenet}.
        To account for class imbalance, we train  weighted cross entropy using inversely proportional class ratios. 
        
        To form prediction sets, we use the nonconformality score described above from~\citet{Sadinle_2018}, which provably has the smallest average set size.
        We calibrate the empirical quantile on the internal validation set.
        We train each model for 50 epochs and average over 5 training runs. 
        
        \begin{figure*}[ht]
        \vskip 0.2in
        \begin{center}
        \centerline{\includegraphics[width=0.9\textwidth]{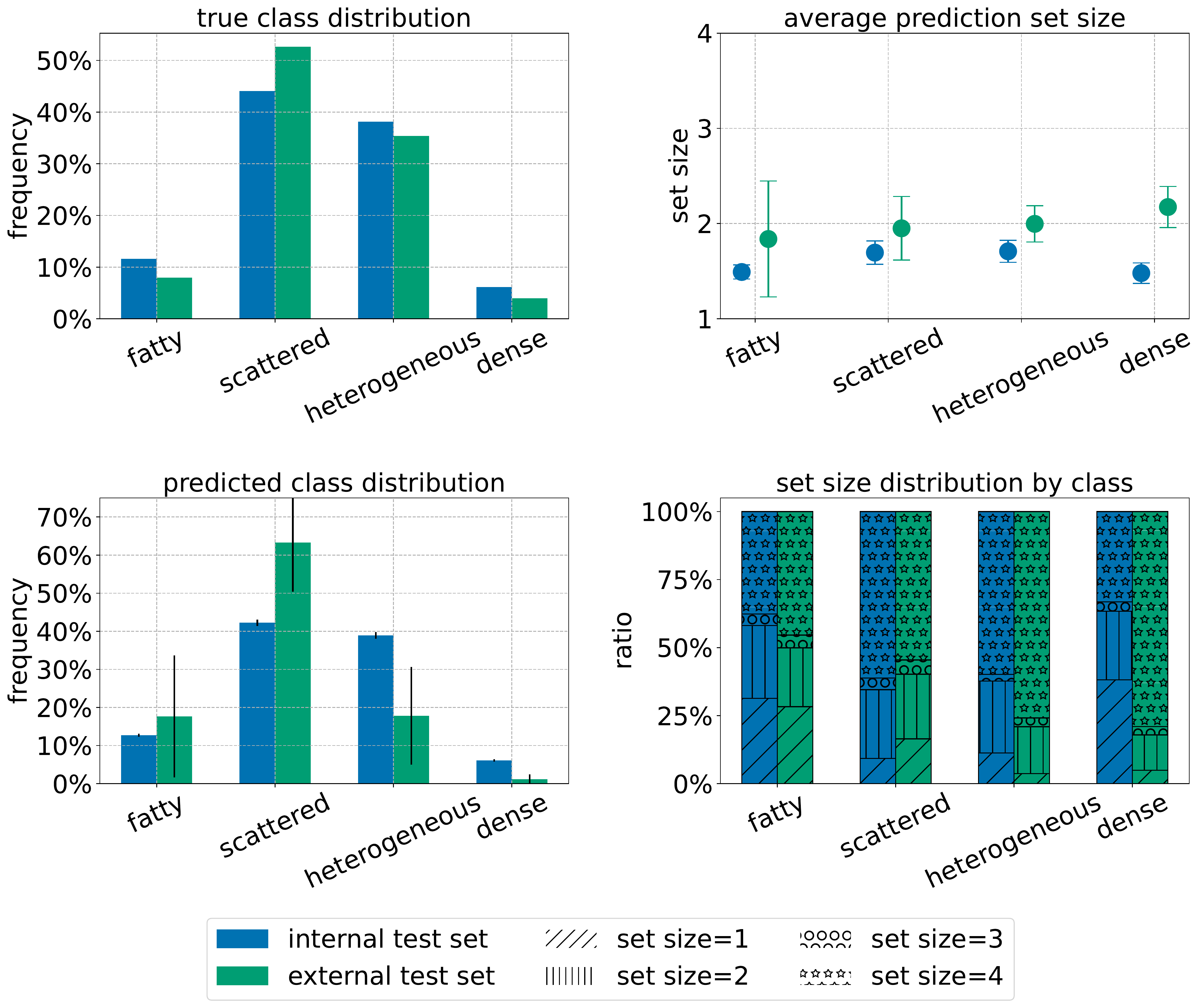}}
        \caption{Comparing internal and external predictions with Efficientnet-B5 model at $\alpha=0.05$ (averaged over 5 runs).}
        \label{fig:compare}
        \end{center}
        \vskip -0.2in
        \end{figure*}
        
        \begin{figure*}[ht]
        \vskip 0.2in
        \begin{center}
        \centerline{\includegraphics[width=0.9\textwidth]{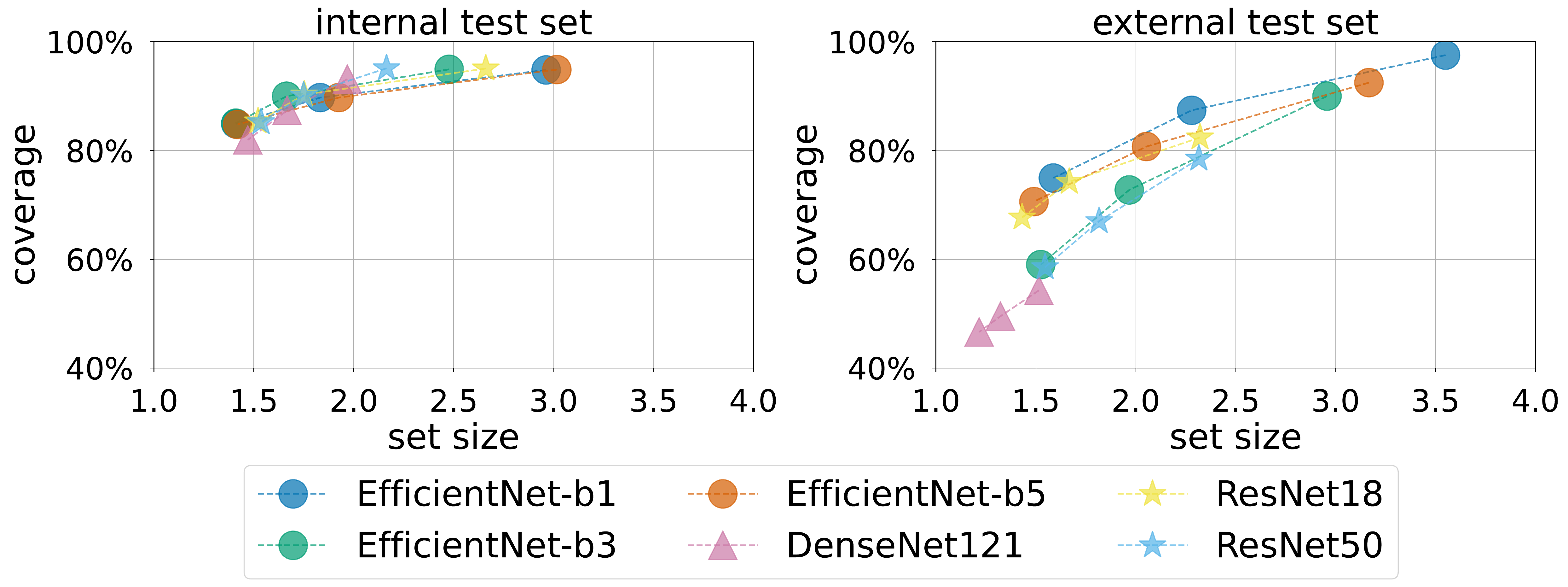}}
        \caption{Comparison of coverage vs set size tradeoff at three different thresholds, $\alpha \in (0.05, 0.1, 0.15)$, for both internal and external test sets (averaged over 5 model runs).}
        \label{fig:arch}
        \end{center}
        \vskip -0.2in
        \end{figure*}
        
        \begin{figure*}[ht]
        \vskip 0.2in
        \begin{center}
        \centerline{\includegraphics[width=0.9\textwidth]{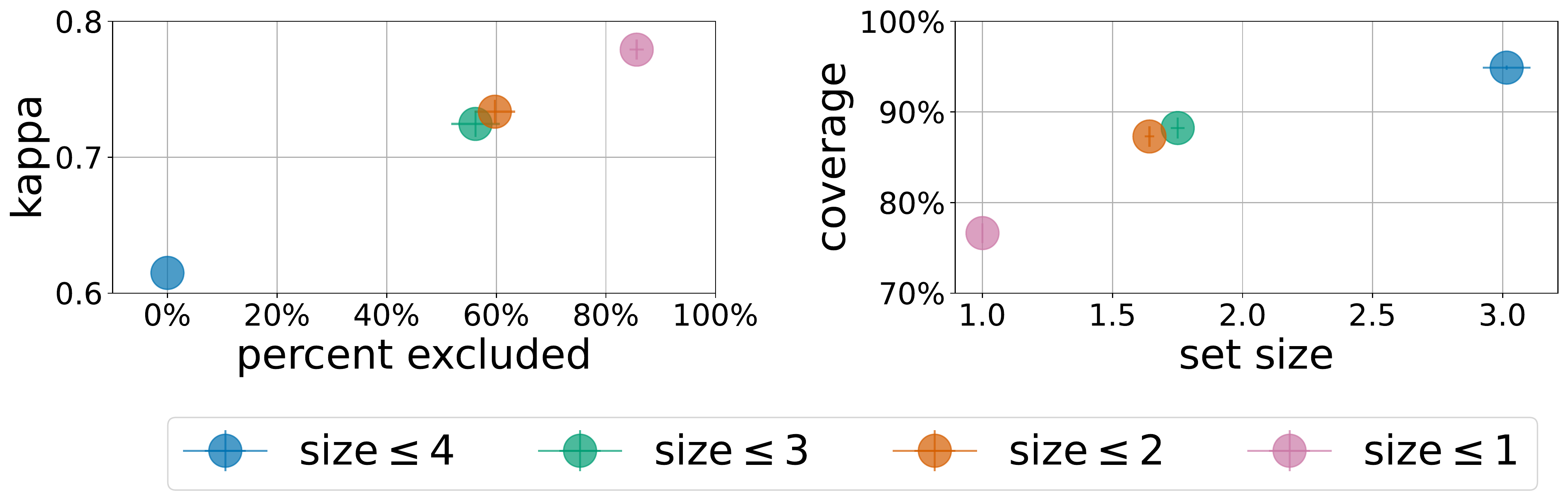}}
        \caption{Filtering out predictions by set size at $\alpha=0.05$ on internal test set with Efficientnet-B5 model, average over 5 runs.}
        \label{fig:filter}
        \end{center}
        \vskip -0.2in
        \end{figure*}
        
        \begin{figure*}[ht]
        \vskip 0.2in
        \begin{center}
        \centerline{\includegraphics[width=0.9\textwidth]{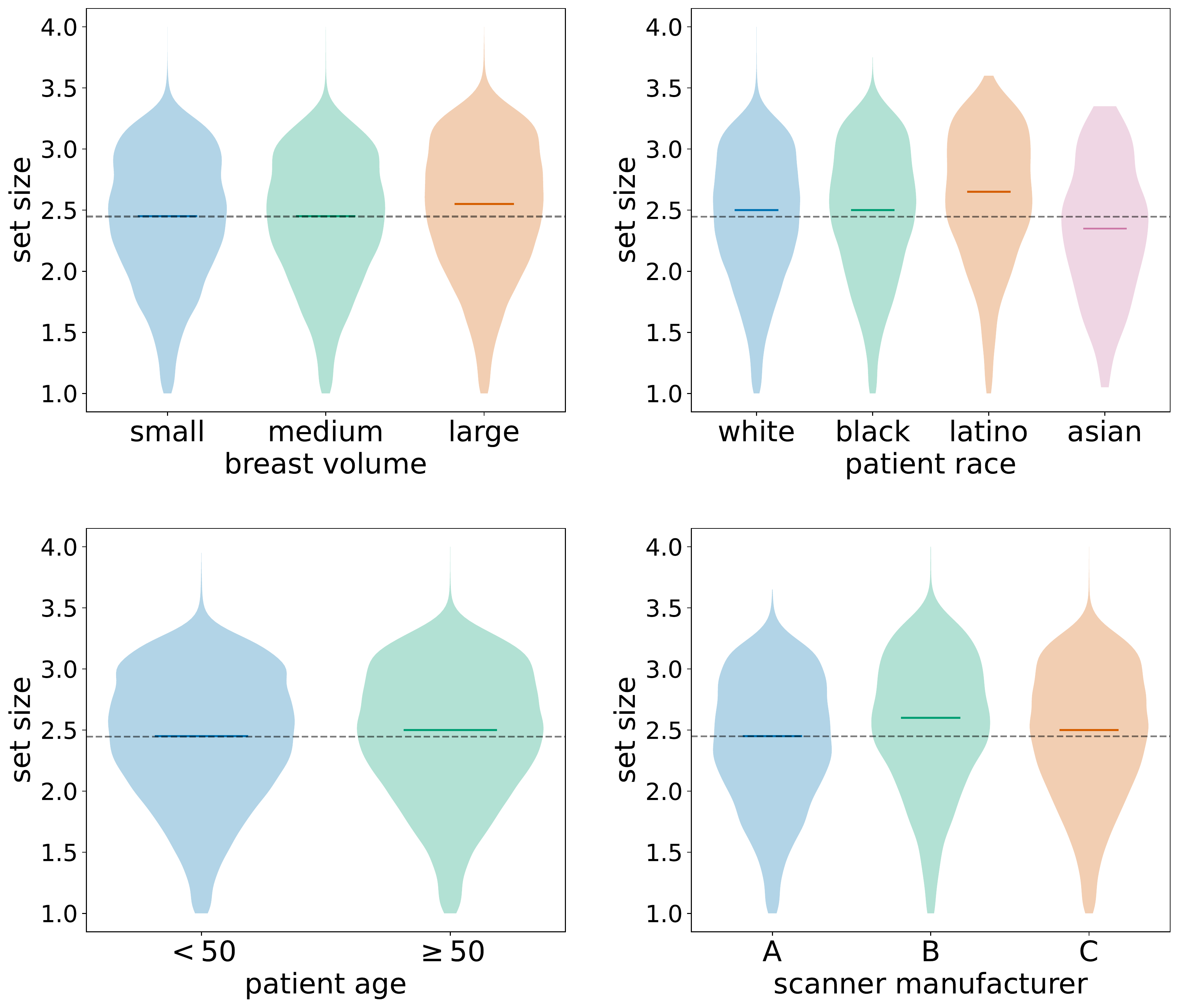}}
        \caption{Violin plots of set size of different cohorts stratified by breast size, patient race, patient age, and scanner type at $\alpha=0.05$ with Efficientnet-B3 averaged over 20 runs.}
        \label{fig:subgroup}
        \end{center}
        \vskip -0.2in
        \end{figure*}

    \subsection{Characterizing Distribution Shift}\label{dist-shift}
        Generalizability to natural distribution shifts is one of the major challenges of machine learning for clinical applications. 
        Models trained and evaluated on one healthcare institution, geographic regions, or scanner type have difficulty maintaining performance to other institutions, regions, and scanners ~\cite{https://doi.org/10.48550/arxiv.2002.11379,10.1001/jama.2020.12067,MARTENSSON2020101714}.
        As CP assumes exchangability in distribution, we would expect most distribution shifts to violate this assumption.
        
        When comparing our internal test set (DMIST) to our external test set (MGH), we observe much higher coverage violations and larger set sizes on the external test set across a variety of architectures, shown in Figure~\ref{fig:arch}.
        
        We can further characterize the differences by examining the distribution of set sizes between internal and external prediction sets in Figure\ref{fig:compare}.
        While the scattered and heterogeneous classes comprise the majority of the breast density ratings, they are more difficult and ambiguous to rate (even for human experts) compared to the fatty and extremely dense classes.
        This inherent uncertainty is reflected in the higher average set sizes for scattered and heterogeneous classes in the internal test set. 
        However in the external test set, the dense class has the highest average set size and both the heterogeneous and dense classes have significantly more sets of size 4 (all classes contained) compared the prediction sets on the internal test set.
        
        This analysis of conformal prediction set provides an alternative insight of distribution shift, that may be helpful in developing techniques and strategies for effective detection and mitigation of different forms of distribution shift.
    
    \subsection{Improving Prediction Quality}\label{filtering}
        Deferring predictions and selective classification can help make deep learning systems more robust in many clinical scenarios where manual review would be preferable to wrong automated predictions~\cite{NEURIPS2018_09d37c08,NIPS2017_4a8423d5}
        
        A second application of CP is the improvement of prediction quality by filtering out low-quality predictions as shown in Figure~\ref{fig:filter}.
        For example, at $alpha=0.05$, we can improve kappa from $0.61$ to $0.72$ by removing the highest uncertainty prediction of set size 4 at a cost of $56\%$ of total predictions and a coverage violation of $0.07$ from the ideal $0.95$.
        
    \subsection{Quantifying Cohort Disparity}\label{disparity}
        Fairness and algorithmic bias has been highlighted as a challenge in medical AI~\cite{DBLP:journals/corr/abs-2003-00827,ahmad2020fairness,gichoya2022ai,https://doi.org/10.48550/arxiv.2110.00603}.
        We believe CP may be highly useful for quantifying disparity between different cohorts to measure efficiency and fairness between clinically relevant demographics.
        
        Using CP calibrated per subgroup, as described in~\citet{https://doi.org/10.48550/arxiv.2109.04392}, we examine the cohorts relevant to breast density: breast volume, patient age, and patient race, in addition to scanner type which has been show to affect performance of deep learning for breast density rating~\cite{DBLP:journals/corr/abs-2103-13511}.
        We observe several statistically significant ($p<0.001$) differences in Figure~\ref{fig:subgroup}, specifically higher average set sizes for large breasts, Latino and/or Hispanic patients, patients older than 50 years of age, and scanner type B.
        Age and breast size are known to be inversely correlated with breast density; however the difference in uncertainty for scanner type B merit additional investigation.
    
\section{Discussion}\label{dicuss}

    Currently, DL algorithms, despite their high promise, has lagged in
translation into clinical practice. 
    One key challenge is the inability to know the trustworthiness of a given model's output. 
    Specifically, whether a prediction for a patient is reliable or should be flagged for further review by experts. 
    Given the high stakes nature of clinical medicine, gaining the trust of the user is crucial. 
    CP methods present a solution by providing a distribution-free measure of uncertainty. 
    Importantly, conformal outputs are highly intuitive to the users while providing theoretical guarantees. 
    Up until this point, conformal predictions have mostly been used in simulated settings. 
    In this paper, we utilize a large multi-institutional medical imaging dataset to show three realistic possible applications. 
    Specifically, we show how CP can be used for three different tasks: distribution shift characterization, prediction quality improvement, and subgroup fairness analysis.

    There are several possible limitations to this study. 
    First, we only 
focus on one medical application within radiology. 
    Further study can evaluate the performance of CP for other diseases and other medical specialities. 

\bibliography{example_paper}
\bibliographystyle{icml2022}

\newpage
\appendix
\onecolumn

\end{document}